\documentclass[prd,floats,aps,epsfig,nofootinbib,axodraw,amssymb,verbatim,preprintnumbers]{revtex4}
\usepackage{amsmath,amsfonts}
\usepackage{subfigure}
\usepackage{verbatim}
\usepackage{slashed}
\usepackage{graphicx}
\usepackage{epstopdf}
\usepackage{color}
\def\beq{\begin{equation}}
\def\eeq{\end{equation}}
\def\bea{\begin{eqnarray}}
\def\eea{\end{eqnarray}} 

\def\Acknowledgements{\bigskip  \bigskip \begin{center} \begin{normalsize}
             \bf ACKNOWLEDGEMENTS \end{normalsize}\end{center}}

\def\leqn#1{(\ref{#1})}

\def\pslash{\not{\hbox{\kern-4pt $p$}}}
\def\qslash{\not{\hbox{\kern-4pt $q$}}}
\def\lv{\not{\hbox{\kern-4pt $L$}}}
\def\lsim{\mathrel{\raise.3ex\hbox{$<$\kern-.75em\lower1ex\hbox{$\sim$}}}}
\def\gsim{\mathrel{\raise.3ex\hbox{$>$\kern-.75em\lower1ex\hbox{$\sim$}}}}
\def\ifmath#1{\relax\ifmmode #1\else $#1$\fi}

\usepackage{graphicx}
\usepackage{bm}
\preprint{SLAC--PUB--16190}
\preprint{MITP/14-104}
\preprint{SCIPP 15/01}

\begin{document}

\title{Perturbative Unitarity Constraints on Charged/Colored Portals}
\bigskip
\author{Matthew Cahill-Rowley$^a$, Sonia El Hedri$^{b}$, William~Shepherd$^c$  and Devin~G.~E.~Walker$^a$}
\address{$^a$SLAC National Accelerator Laboratory, 2575 Sand Hill Road, Menlo Park, CA 94025, U.S.A.}
\address{$^b$PRISMA Cluster of Excellence and Mainz Institute for Theoretical Physics Johannes Gutenberg
University, 55099 Mainz, Germany}
\address{$^c$Santa Cruz Institute for Particle Physics and Department of Physics, Santa Cruz, CA 95064, U.S.A.}

\begin{abstract}
\noindent
Dark matter that was once in thermal equilibrium with the Standard Model is generally prohibited from obtaining all of its mass from the electroweak or QCD phase transitions.  This implies a new scale of physics and mediator particles needed to facilitate dark matter annihilations.  In this work, we consider scenarios where thermal dark matter annihilates via scalar mediators that are colored and/or electrically charged.  We show how partial wave unitarity places upper bounds on the masses and couplings on both the dark matter and mediators.  To do this, we employ effective field theories with dark matter as well as three 
flavors of sleptons or squarks with minimum flavor violation.   For Dirac (Majorana) dark matter that annihilates via mediators charged as left-handed sleptons, we find an upper bound around $45$~TeV ($7$~TeV) for both the mediator and dark matter masses, respectively.  These bounds vary as the square root of the number of colors times the number of flavors involved. Therefore the bounds diminish by root two for right handed selectrons.  The bounds increase by root three and root six for right and left handed squarks, respectively.  Finally, because of the interest in natural models, we also focus on an effective field theory with only stops.  We find an upper bound around $32$ TeV (5 TeV) for both the Dirac (Majorana) dark matter and stop masses.  In comparison to traditional naturalness arguments, the stop bound gives a firmer, alternative expectation on when new physics will appear.  Similar to naturalness, all of the bounds quoted above are valid outside of a defined fine-tuned regions where the dark matter can co-annihilate.  The bounds in this region of parameter space can exceed the well-known bounds from Griest and Kamionkowski~\cite{Griest:1989wd}.  We briefly describe the impact on planned and existing direct detection experiments and colliders.  
\end{abstract}
\maketitle

\section{Introduction}
\label{sec:intro}
\noindent
Understanding the nature of dark matter is one of the most pressing, unresolved problems in particle physics.  Dark matter is needed to understand structure formation, the observed galactic rotation curves~\cite{Kowalski:2008ez,Ahn:2013gms,Beringer:1900zz} and the acoustic peaks in the cosmic microwave background~\cite{Ade:2013lta}.  Moreover, the dark matter relic abundance is measured to be~\cite{Ade:2013lta}  
\begin{equation}
h^2\, \Omega_c = 0.1199 \pm 0.0027.  \label{eq:relic}
\end{equation}
A compelling argument for the origin of this abundance is to assume dark matter was once in thermal contact with the baryon-photon plasma during the early universe.  Since all known forms of matter in the universe were once in thermal equilibrium, this type of dark matter is theoretically persuasive.  In this scenario, the measured relic abundance is controlled by dark matter annihilations into Standard Model (SM) particles.  Because of constraints from the observed large scale structure in the universe, dark matter must be stable and non-relativistic (cold) when departing thermal equilibrium~\cite{Ahn:2013gms}.  
\newline
\newline
The Standard Model (SM) alone cannot account for the missing matter in the universe~\cite{Bertone:2004pz}.  Current experimental constraints, however, provide some guidance on the structure of the underlying theory.  For example, the lack of large missing energy signatures at the Large Hadron Collider (LHC)~\cite{Aad:2014tda,ATLAS:2014wra,Aad:2013oja,Aad:2014vka,Aad:2014kra,Chatrchyan:2014lfa,Khachatryan:2014tva,Khachatryan:2014rwa,Khachatryan:2014rra,Chatrchyan:2013lya,Chatrchyan:2013xna} and other colliders~\cite{Aaltonen:2012jb,Aaltonen:2013har,CDF:2011ah,Abachi:1996dc,Abazov:2012qka,Abdallah:2003np,Acciarri:1997dq,Abbiendi:2004gf} suggests that the dark matter is either heavy or has very small couplings with the SM so that it is not produced in high-energy collisions.  Additionally, direct detection experiments~\cite{Aprile:2012nq,Ahmed:2011gh,Akerib:2013tjd}, updated precision electroweak constraints, and precision Z-pole experiments~\cite{devinjoannetim,Baak:2012kk,ALEPH:2005ab} all severely constrain the direct coupling of dark matter to the SM Higgs and/or Z~bosons.  These constraints all imply that dark matter cannot obtain all of its mass from the SM Higgs alone~\cite{devinjoannetim}.  Mediator-facilitated interactions help to evade current experimental constraints by partially decoupling the dark matter from the SM.  Should these scenarios be realized in nature, the discovery of the mediator particles would be an important step in understanding the nature of dark matter.  It is therefore crucial to place bounds on the masses and couplings of these mediators.  %
The most popular ways for dark matter to annihilate via a mediator particle are through scalars that are colored or charged, the Higgs boson~\cite{Patt:2006fw},  or via a new neutral gauge boson.  Some of us explored the perturbative unitarity constraints on the Higgs~\cite{Walker:2013hka,Betre:2014sra,Betre:2014fva} and the gauge portals~\cite{Hedri:2014mua}.  In this work we focus on constraining scenarios where fermionic dark matter annihilates via charged or colored scalars.  
\newline
\newline
In supersymmetric (SUSY) models, the charged and colored scalars are sleptons and squarks, respectively.  
We employ an effective field theory including these particles as a benchmark in our analysis.  %
%
In addition, because of the intense focus on natural SUSY models, we particularly consider an effective field theory with dark matter and %
the supersymmetric partner of the top quark (stop).  The stop is the key for achieving natural, supersymmetric models of new physics.  Note naturalness~\cite{Susskind:1978ms} has been the dominant paradigm that has shaped models of new physics for the last 35 years.  It is minimally satisfied with a relatively light stop~\cite{Cohen:1996vb} which cancels the quadratic divergences to the SM Higgs mass from the top quark.  Moreover if the neutralino dark matter is only \textit{one component} of the measured relic abundance (e.g., axionic plus neutralino dark matter), then the constraint on the stop mass improves.  Thus, our bounds may be useful in shaping viable R-parity conserving supersymmetric models. 
\newline
\newline
We apply unitarity constraints in a manner reminiscent of Griest and Kamiokowski~\cite{Griest:1989wd}.  However, there are important differences:  here we focus on perturbative unitarity constraints which determine, in particular, when the dark matter couplings become strong.  Importantly, weakly interacting, massive dark matter 
and perturbativity have always had an intertwined conceptual association.  %
In models that violate perturbative unitarity, the elementary dark matter candidate is efficiently forming bound states as well as annihilating as the temperature decreases towards the thermal decoupling temperature.  The annihilation diagrams for the elementary dark matter candidate can be easily altered (and often times dressed with lighter states) to form diagrams in which the bound state decays.  The bound state will therefore have a lifetime much shorter than the age of the universe.  
Thus, in models that violate perturbative unitarity, the elementary dark matter candidate is not an asymptotic state and the bound state that it forms is not a viable dark matter candidate.  Note, it is well known that viable dark matter candidates exist that are the result of strongly coupled or confining hidden sectors. However, in these models the dark matter annihilation processes are still perturbative~\cite{Shepherd:2009sa}.  We show our perturbative unitarity constraints are improved in comparison to the updated Griest and Kamiokowski bounds~\cite{Profumo:2013yn}.  Of central importance is the fact that our methodology places constraints on any particle associated with the dark matter annihilation.  Our bounds on the masses and couplings of the charged and colored scalars are novel.  %
\newline
\newline
Our basic perturbative unitarity arguments are straightforward.  Dark matter annihilating via the electrically charged and/or colored scalars does so via t-channel diagrams.  As one increases the mass of the mediator, the annihilation cross section is diminished unless the Yukawa coupling (between the fermionic dark matter, SM fermions and mediators) is raised.  A sufficiently large annihilation cross section is needed to satisfy the relic abundance constraint.  This process also forces the dark matter mass to be raised as well.  Eventually, the mediator mass is sufficiently large that the couplings violate perturbative unitarity therefore setting an upper bound on both the dark matter and mediator masses.  The parameter space for this process is additionally constrained by requiring the dark matter relic abundance to be less than or equal to equation~\leqn{eq:relic}.  Conversely, small mediator masses are constrained by collider searches and other measurements.
\newline
\newline
In the next section, we introduce simple models onto which we place our unitary and relic abundance bounds.  Section III.~introduces our unitarity constraints.  Section IV.~computes the dark matter relic abundance and direct detection cross sections.  Section V.~describes existing experimental constraints from the LHC.  We implement all of our constraints and perform a parameter scan in Section VI.  The basic unitarity bounds appear here.  Future experimental signatures are in Section VII.  Conclusions and Appendices follow.
\section{Representative Sfermion Models}
\label{Sec: sec2}
%
\noindent
Throughout we consider effective theories where the only relevant degrees of freedom are a SM singlet fermionic dark matter particle (denoted by $\chi$ and potentially having either a Dirac or Majorana mass), a scalar mediator, and the SM particles.  %
For this class of models, we consider scalar mediators that facilitates the following tree-level DM annihilation processes,
\begin{align}
\bar{\chi} + \chi \to \bar{f} + f, 
\label{eq:annihilation}
\end{align}
through a Yukawa-type coupling between the SM fermion $f$, the DM particle $\chi$, and the scalar mediator.  As a result, $f$ must have the same SM quantum numbers as the scalar mediator.  Because we consider all possible SM final states, we consider scenarios with all possible SM quantum numbers for the scalar mediators.  %
In order to stabilize the DM, both it and the scalar mediator must be charged under the same dark matter stabilization symmetry.  Because our scalar mediators are similar to the sfermions in SUSY models, we will therefore employ the SUSY nomenclature to describe them. However, the key difference with respect to SUSY theories is that we allow the DM-sfermion-fermion coupling to take an arbitrary value up to the perturbative limit.
\newline
\newline
We consider an effective scenario in which the DM-sfermion-fermion coupling is the only non-gauge interaction for the sfermion~\cite{DiFranzo:2013vra}.  The sfermions will either be purely left-handed, $\tilde{f}_{L}$, or right-handed states $\tilde{f}_{R}$. Calling the DM-sfermion-fermion coupling $\lambda_i$, and considering a single set of sfermion quantum numbers, the interaction Lagrangian is  
\begin{equation}
\mathcal{L} =  \mathcal{L}_\mathrm{gauge} +  \mathcal{L}_\mathrm{mediators} + \mathcal{L}_\chi 
\label{Eq: lag1}
\end{equation}
where
\begin{align}
\mathcal{L}_\mathrm{gauge} = - {1 \over 4} G_{\mu\nu}\, G^{\mu\nu}  - {1 \over 4} F_{\mu\nu}\,F^{\mu\nu} &&
\mathcal{L}_\chi  =   i \bar{\chi} \partial\!\!\!\slash \chi - m_\chi \bar{\chi}\chi \label{Eq: lag2}
\end{align}
and the field strength tensors for the gluon and photon are $G_{\mu\nu}$ and $F_{\mu\nu}$, respectively.  Also, depending on the effective scenario the mediator lagrangian is 
\begin{align}
\mathcal{L}_\mathrm{mediators} = \bigl(D \tilde{f}_{i}\bigr)^* D\tilde{f}_{i} - V_i 
\end{align}  
where $i = L, R$  and the covariant derivative, $D$, depends on the quantum numbers of $\tilde{f}_{L}$ or $\tilde{f}_{R}$.  The potential is 
 \begin{align}
 V_i = \frac{1}{2} m_{\tilde{f} i}^2 \,\tilde{f}_{i}^* \tilde{f}_{i}  + \Bigl( \lambda_i \,\tilde{f}^*\, \bar\chi\,P_{i} \,f + \mathrm{h.c.}\Bigr) 
\label{Eq: lag3} 
 \end{align}
We remain agnostic about the mechanism that generates the $\chi$ mass and assume it can be integrated out.  %
Because we chose  the Yukawa-type coupling between the dark matter, mediator and fermion to be non-zero, our analysis is generic and the derived bounds subsequently weaker in comparison to~\cite{deSimone:2014pda} which considers only the QCD gauge coupling between those particles.  Once the quantum numbers of the sfermion are chosen, our scenario then has three basic independent parameters,
\begin{align}
\left\{\lambda, m_{\tilde{f} i}, m_\chi\right\}.
\end{align}
These parameters are constrained by direct detection experiments, collider searches (see the relevant direct detection and collider citations in the Introduction) and unitarity constraints.  Since we assume thermal dark matter and require $\chi$ to comprise some or all of the observed DM in the universe, our effective models are also constrained by the DM relic abundance.  %
\newline
\newline
We consider scenarios with mediators that are color singlets as well as color triplets.  The mediators may also be electroweak doublets as well.   Although the perturbative unitarity bounds do not depend on the sfermion quantum numbers, the effects of these quantum numbers on the relic density can be reproduced through the following rescaling %
%
%
\begin{equation}
 \lambda\tilde{f}^*\, \bar\chi\,P_{i} \,f  \to  N^{1/4}\,\lambda\,\tilde{f}^*\, \bar\chi\,P_{i} \,f 
\end{equation} 
where the rescaling factor $N$ depends on the representations of the groups $\tilde f_i$ is charged under. More details about how to compute $N$ are given in Sec.~\ref{Sec: sec4}. Here $N$ encodes the dimensions of the representations of the gauge group under which the mediator and SM fermions are charged.  As $N$ is increased, in general larger dark matter and mediator masses are possible because the dark matter annihilation is more efficient.  Thus, there are weaker constraints from the relic abundance.  As we show in Section~\ref{Sec: sec4}, our basic bounds scale accordingly.  We briefly describe each scenario below.  %
%
%
%

\subsection{Colored Flavor Triplet Effective Model} 
\noindent
The colored mediators can have the quantum numbers 
\begin{align}
(3,2)_{1/6} && (\bar{3},1)_{-2/3} && (\bar{3},1)_{1/3} .
\end{align}
We consider all three possible types of mediators separately and will call them $\tilde q_L$, $\tilde{u}_R$ and $\tilde{d}_R$ respectively. In order to prevent flavor violation, we will consider the case in which these mediators are degenerate flavor triplets,
\begin{align}
\tilde{q}_L &= 
\begin{pmatrix}
\tilde u_L\\
\tilde d_L
\end{pmatrix}\quad 
\begin{pmatrix}
\tilde c_L\\
\tilde s_L
\end{pmatrix}\quad
\begin{pmatrix}
\tilde t_L\\
\tilde b_L
\end{pmatrix}&&
\tilde u_R = 
\begin{pmatrix}
\tilde t_R\\
\tilde c_R\\
\tilde u_R
\end{pmatrix}&&
\tilde d_R = 
\begin{pmatrix}
\tilde b_R\\
\tilde s_R\\
\tilde d_R
\end{pmatrix}.
\end{align}
with the DM-sfermion-fermion coupling matrix $\lambda$ proportional to the identity matrix. The collider, relic and direct detection constraints (but not the perturbative unitarity constraints) have been previously studied in~\cite{DiFranzo:2013vra}.
\newline
\newline
To make our work more comprehensive, in the Appendix we modify the above scenario to allow for non-trivial couplings between the squark mediators and the SM Higgs.  These couplings are analogous to the A-terms in SUSY models.   Thus after electroweak symmetry breaking (EWSB), left-right squark mixing is generated.  
We focus on a model with one left-handed squark doublet $\tilde{q}_L$ and one right-handed up-type squark $\tilde{u}_R$, with the right-handed down-type squark $\tilde{d}_R$ decoupled.  Of note, this scenario has also been studied in the context of SUSY for the squark-mediated annihilation of bino-like dark matter~\cite{Kelso:2014qja}. 

\subsection{Color Singlet Flavor Triplets Effective Model}
\noindent
In this effective scenario, the color singlet mediators must have the quantum numbers
\begin{align}
(1,2)_{-1/2} && (1,1)_1.
\end{align}
We denote these mediators as $\tilde{l}_L$ and $\tilde{e}_R$ and call them sleptons.  Once again, we consider them in degenerate flavor triplets,
\begin{align}
\tilde{l} = \begin{pmatrix}
\tilde e\\
\tilde \nu_e
\end{pmatrix}\quad
\begin{pmatrix}
\tilde e\mu\\
\tilde \nu_\mu
\end{pmatrix}\quad
\begin{pmatrix}
\tilde \tau\\
\tilde \nu_\tau
\end{pmatrix}&&
\tilde e_R = \begin{pmatrix}
\tilde \tau\\
\tilde \mu\\
\tilde e
\end{pmatrix}.
\end{align}
with $\lambda$ again proportional to the identity. The annihilation cross-section for color singlet mediators is decreased with respect to the colored mediator cross section.  
Thus, the relic density will play a larger role in constraining this scenario. Direct detection experiments will not be able to constrain this scenario at tree-level.

\subsection{Stop Effective Model}
 %
%
\noindent
Natural SUSY requires, at minimum, a light stop to cancel the SM Higgs mass quadratic divergences generated by SM top loops~\cite{Cohen:1996vb}.  Thus far, the LHC has not discovered the stop, and is setting increasingly stringent constraints on the stop mass~\cite{Aad:2014kra,Chatrchyan:2013xna}.  Thus, because of its significance to natural theories of new physics, we also consider the case where the right-handed (RH) stop, $\tilde{t}_R$, is much lighter than the other squark flavors.  Although we only focus on light RH stop, models where only the left-handed (LH) stops is light should be constrained in a similar way.  The LHC and, if necessary, a future hadronic collider is essential to discover/constrain this scenario.  
\section{Unitarity Constraints}
\label{Sec: sec3}
\begin{figure}
\centering
\includegraphics[width=0.4\linewidth]{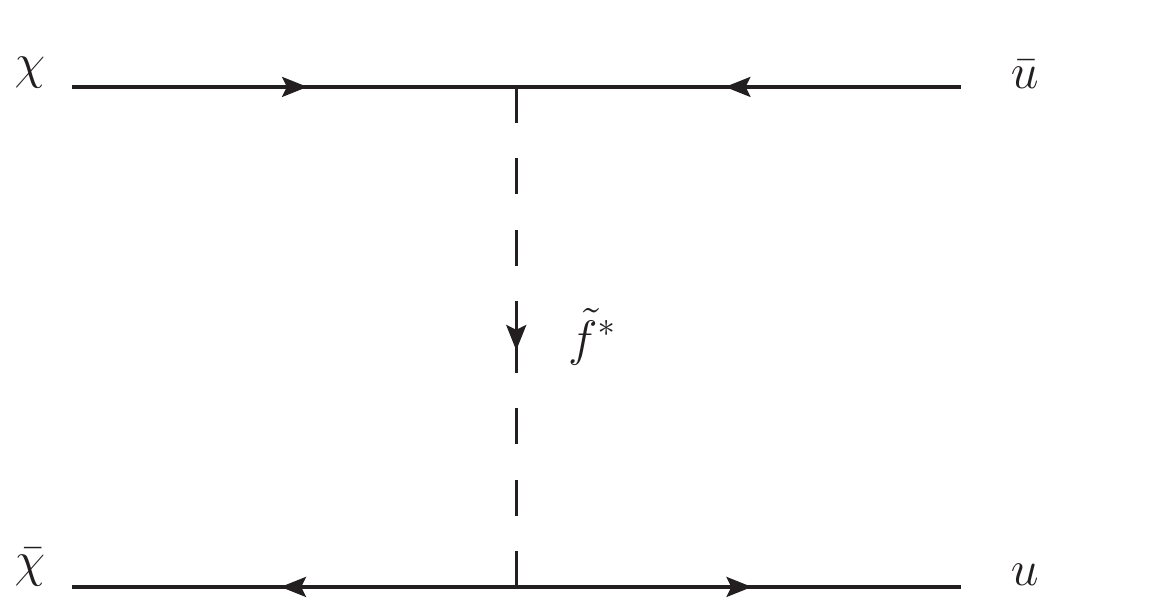}\\
\includegraphics[width=0.4\linewidth]{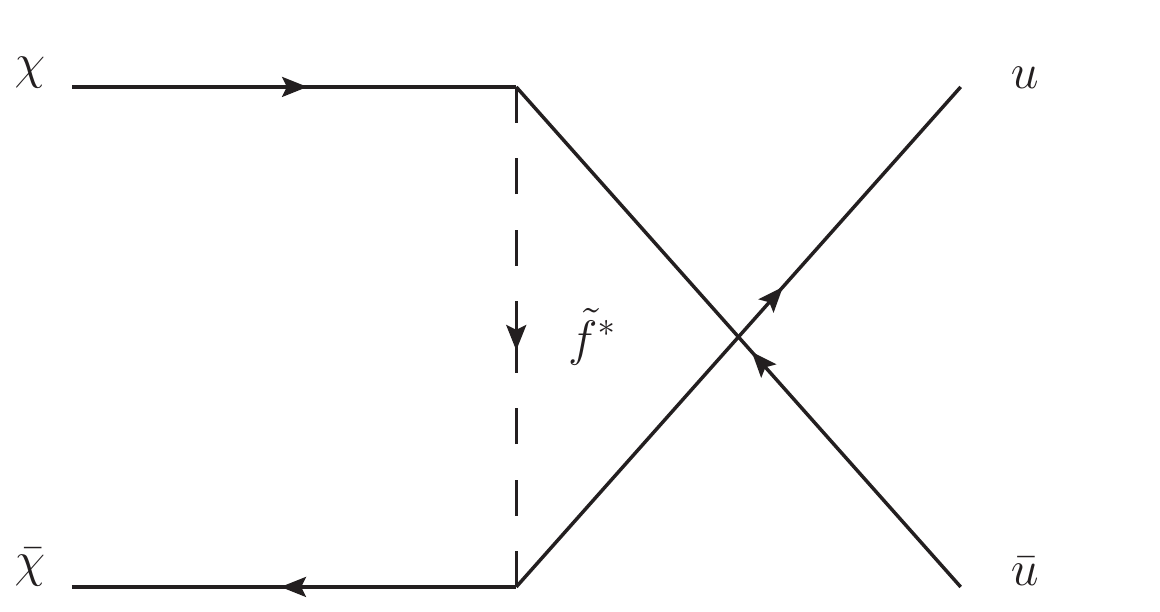}
\caption{\label{Fig: DMunitarity} Diagrams contributing to the $\bar \chi \chi \rightarrow \bar f f$ amplitude for Dirac and Majorana DM. For Dirac DM, only the $t$-channel (top) diagram contributes to the scattering amplitude.}
\end{figure}

\noindent
We consider an effective theory where a charged/colored scalar (sfermion) mediates dark matter annihilation.   The only couplings allowed to vary freely are the DM-fermion-sfermion Yukawa couplings, $\lambda_{i}$ where $i = L, R$ from equation~\leqn{Eq: lag3}.  These couplings drive all the DM scattering and annihilation processes and are directly constrained by perturbative unitarity.  As discussed in the Introduction, there has been a long understood conceptual dependence of the WIMP dark matter paradigm on perturbativity.  
%
\newline
\newline
It is well known~\cite{Aydemir:2012nz,Schuessler:2007av,Schuessler:thesis} that any theory known to all orders in perturbation theory gives amplitudes which lie on the Argand circle.  %
%
However most perturbative unitarity computations are done at tree-level and the loop corrections push the value of the scattering amplitudes toward the Argand circle.  %
%
%
If the scattering amplitudes of some processes in the theory are too large at tree-level, unitarity cannot be restored without violating perturbativity. \cite{Schuessler:2007av,Schuessler:thesis,Betre:2014fva} have derived a simple geometric argument that allows the translation of a perturbativity bound into an upper bound on the tree-level scattering amplitudes in any theory. This upper bound applies to the eigenvalues of the partial wave components of the transition operator $T$. This operator is defined such that
\begin{align}
S = 1 + i \,T
\end{align}
where $S$ is the S-matrix.  In this work, we require that the loop contributions to the scattering amplitudes amount to less than $40\%$ of the tree-level values. According to~\cite{Schuessler:2007av,Schuessler:thesis}, this amounts to enforcing
\begin{align}
\left|\mathcal{T}^j_{ii}\right| < \frac{1}{2}
\label{Eq: unitarity}
\end{align}
where $\mathcal{T}^j_{ii}$ is the $j$-th partial wave component of $T$. Since the lowest modes generally give the strongest constraints, we only focus on the s-wave component of $T$ ($j = 0$). 
\newline
\newline
In the following, we derive unitarity bounds for the scattering of the $\bar \chi \chi$ and $\bar f f$ pairs where $f$ is any of the fermions associated to the scalars decribed in Section~\ref{Sec: sec2}. We compute the bounds in the large c.o.m. energy limit
\begin{align}
s \gg M_\chi, \,M_{\tilde f}, \,v,
\end{align}
where $v$ is the electroweak vev.  $\bar \chi \chi \rightarrow \bar\chi \chi$ scattering is forbidden at tree level in our model. At the center of mass energies considered here, both the electroweak and strong couplings can be neglected compared to $\lambda$. The amplitude associated to $\bar f f \rightarrow \bar f f$ can therefore be neglected. We are thus left with only one process to consider,
\begin{align}
\bar \chi \chi \rightarrow \bar f f,
\end{align}
which is entirely determined by $\lambda_L$ and/or $\lambda_R$. %
If the DM is a Dirac particle, $\bar \chi \chi \rightarrow \bar f f$ can only occur in the t-channel, through the diagram shown in Figure~\ref{Fig: DMunitarity}. Using the results quoted in \cite{Chanowitz:1978uj,Chanowitz:1978mv}, the amplitude for this diagram in the large $s$ limit is 
\begin{align}
\mathcal{M} &= -\lambda^{(i)2} \delta(\alpha_\chi, \alpha_f)\delta(\alpha_{\bar\chi}, \alpha_{\bar f})
\end{align}
where $\lambda^{(i)}$ is either $\lambda_L$ or $\lambda_R$, $\delta$ is the Kronecker symbol and $\alpha_f$ is the sign of the helicity of a given fermion $f$. Here, in the large $s$ limit, we denote the helicity of the LH fermions as $-$ and the one of the RH fermions as $+$. Since we work at large $s$, we can operate in the interaction eigenstate basis. For generic sfermions $\tilde f_R, \tilde f_L$, we then consider the following fermion pairs
\begin{align}
\overline{\chi}_L\chi_L,\, \overline{\chi}_R\chi_R,\, \overline{f}_Lf_L, \,\overline{f}_R f_R, 
\end{align}
and find
\begin{align}
\mathcal{T}^0 &=\frac{1}{16\pi}
\begin{pmatrix}
0 & 0 & 0 & \lambda_R^2\\
0 & 0 & \lambda_L^2 & 0\\
0 & \lambda_L^2 & 0 & 0\\
\lambda_R^2 & 0 & 0 & 0
\end{pmatrix}.
\end{align}
\noindent
The eigenvalues of this matrix are $
\pm \lambda_R^2$ and  $\pm\lambda_L^2$.
For $\theta\in \left[0, \frac{\pi}{4}\right]$, the unitarity bound equation~\leqn{Eq: unitarity} then becomes
\begin{align}
\left|\lambda_L\right|, \left|\lambda_R\right| < \sqrt{8\pi}\sim 5.
\end{align}
In the case of Majorana DM, both the t and u-channel diagrams contribute to the $\bar \chi \chi \rightarrow \bar f f$ process (see Figure~\ref{Fig: DMunitarity}). These diagrams interfere constructively and the resulting amplitude is doubled with respect to the Dirac case. The bounds quoted above then become
\begin{align}
\left|\lambda_L\right|, \left|\lambda_R\right| < \sqrt{4\pi}\sim 3.5.
\end{align}
  \section{Dark Matter Constraints}
\label{Sec: sec4}
\noindent
The DM constraints are effectively determined by the coupling $\lambda_i$ associated to the DM-fermion-sfermion vertex.  See equation~\leqn{Eq: lag3}.  This vertex parameterizes DM annihilation and, for light flavor squarks, DM-nucleus scattering at tree-level.  
 
\subsection{Dark Matter Relic Abundance}
\begin{figure}
\centering
\includegraphics[width=0.6\linewidth]{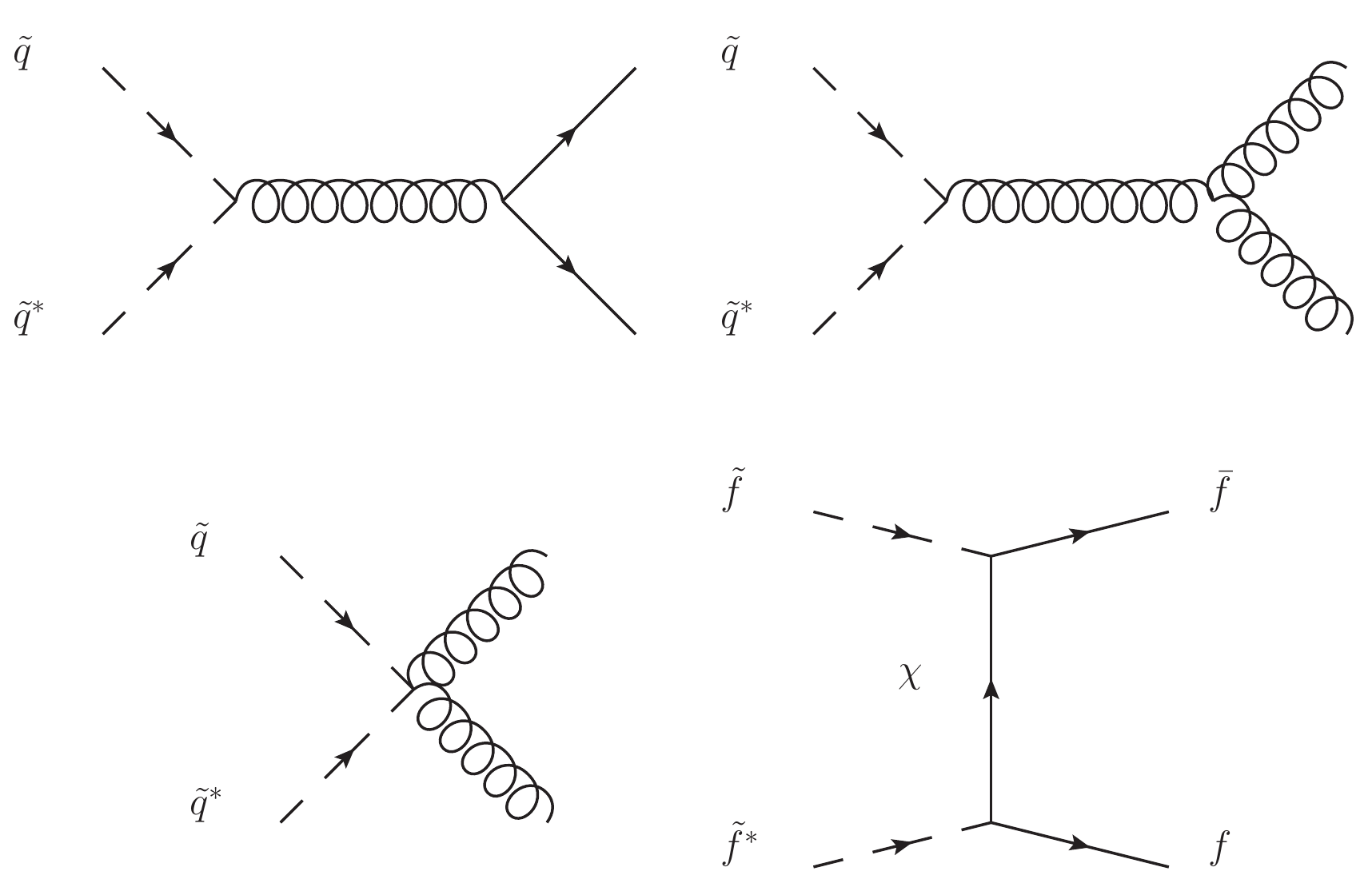}\\
\bigskip
\includegraphics[width=0.35\linewidth]{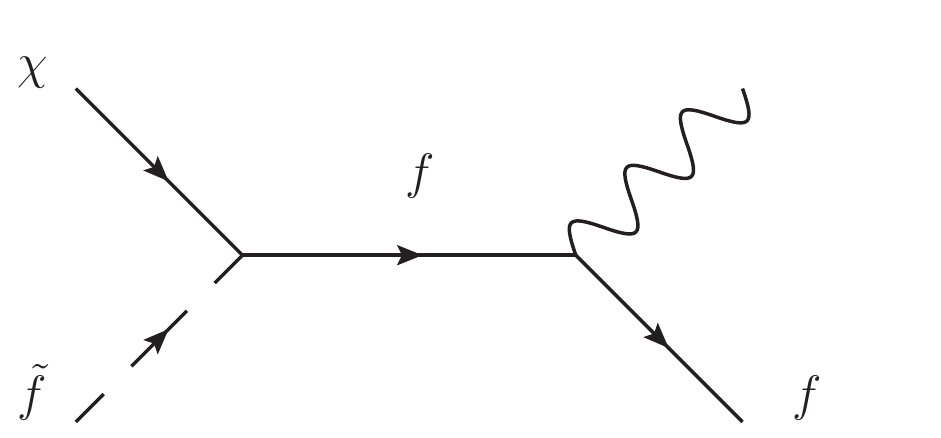}
\caption{\label{Fig: coannihilation}Example of coannihilation diagrams for the dark matter $\chi$ and the sfermions $\tilde f$. Here, the wavy lines represent any gauge boson, including the gluon.}
\end{figure}

\noindent
Dark Matter can annihilate to fermions via the diagrams shown in Figure~\ref{Fig: DMunitarity} for Dirac and Majorana DM. For Dirac DM, annihilation occurs dominantly in the s-wave, with the s-wave term $a$ being
\begin{align}
a\propto N_C N_{\mathrm{flavors}}\,\frac{\lambda^4}{8\pi}\frac{m_\chi^2}{(m_\chi^2 - m_{\tilde f}^2)^2}
\end{align}
where $N_C$ is the color factor. 
When the DM and sfermion masses increase, the coupling $\lambda$ also needs to increase to ensure efficient DM annihilation, unless the DM and sfermion masses are nearly degenerate. The coannihilation region, where the DM and mediator masses are within a few percent of each other, is however extremely narrow and can only be obtained at the price of a significant fine-tuning. Throughout this paper, we will define this coannihilation region as the region where the DM and sfermion masses are within at most $5\%$  of each other.  Outside the coannihilation region, the s-wave component of the DM annihilation cross section is bounded by
\begin{align}
a_{\mathrm{max}} &\simeq 100 \,N_C \,N_{\mathrm{flavors}}\,\frac{\lambda^4}{m_\chi^2}.
\end{align}
In a perturbative theory, unitarity sets an upper bound on $\lambda$, as studied in Section~\ref{Sec: sec3}. When $\lambda$ reaches its maximum value, $a_{\mathrm{max}}$ will decrease as the Dark Matter mass increases. The unitarity constraints on $\lambda$ therefore translate into an upper bound on $m_\chi$. This upper bound scales as $\sqrt{N_C N_{\mathrm{flavors}}}$, independently on the other properties of the mediators involved.  A similar reasoning leads to a neighboring upper bound on the mediator masses.
\newline
\newline
A similar mechanism is at play for Majorana DM. In this case, however, DM annihilates through both the $t$ and the $u$-channels, as shown in Figure~\ref{Fig: DMunitarity}. The two corresponding diagrams interfere destructively.  The s-wave component of the DM annihilation cross section to two fermions $f$ is therefore suppressed by a factor of $m_f^2$.  Heavy DM annihilation then occurs mostly in the p-wave \cite{Jungman:1995df}. 
As computed in \cite{Chang:2013oia}, the annihilation cross section for Majorana DM goes as
\begin{align}
\sigma_{\mathrm{ann}} &\sim  \sum_i\lambda^4\left(\frac{m_f^2}{m_{\tilde f_i}^4} + v^2 \frac{m_\chi^2}{m_{\tilde f_i}^4}\right).
\end{align}
 In the early universe, the velocity suppressed p-wave term dominates and, for degenerate sfermion masses, all the DM annihilation channels contribute equally to the relic density. This cross section is in general much smaller than the annihilation cross section for Dirac DM. Outside the coannihilation region, the relic density constraints on Majorana DM are then expected to be significantly tighter than the constraints on Dirac DM.
\newline
\newline
Inside the coannihilation region, Dark Matter can coannihilate with the sfermions through the diagrams shown in Figure~\ref{Fig: coannihilation} \cite{Griest:1990kh,Edsjo:1997bg,Ellis:1998kh,Edsjo:2003us}. This process can significantly decrease the DM relic density and will considerably loosen the bounds obtained through the method described above. Note that the Griest and Kamionkowski bound \cite{Griest:1989wd} does not take coannihilation into account so, in this region, DM can be much heavier than $120$ $\mathrm{TeV}$ without overclosing.


\subsection{Direct Detection}
\noindent
DM-nucleus scattering occurs in the $s$ and $t$ (and $u$ for Majorana DM) channels through intermediate squarks. In this work, we consider only tree-level DM-nucleus scattering and therefore focus on the light flavor squark components of $\tilde u_R$ $\tilde d_R$ and $\tilde q_L$. The spin-independent (SI) direct detection (DD) cross sections for Dirac DM for $\tilde u_R$ $\tilde d_R$ and $\tilde q_L$ have been computed in \cite{DiFranzo:2013vra} and are approximately
\begin{align}
\sigma_{SI}^{\tilde u_R} &\sim \frac{\mu^2}{64\pi}\frac{\lambda^4}{(m_\chi^2 - m_{\tilde u_R}^2)^2}\left(1 + \frac{Z}{A}\right)^2\\
\sigma_{SI}^{\tilde d_R} &\sim \frac{\mu^2}{64\pi}\frac{\lambda^4}{(m_\chi^2 - m_{\tilde d_R}^2)^2}\left(2 - \frac{Z}{A}\right)^2\\
\sigma_{SI}^{\tilde q_L} &\sim \frac{9\mu^2}{64\pi}\frac{\lambda^4}{(m_\chi^2 - m_{\tilde q_L}^2)^2}.
\end{align}
Constraints from DD are in tension with relic density constraints since they prevent $\lambda$ from being too large. 
\newline
\newline
For Majorana DM, the operator responsible for spin-independent DM-nucleon scattering is generated at higher order in the effective operator expansion \cite{Chang:2013oia,An:2013xka,Garny:2014waa,Drees:1993bu}. Outside the coannihilation region, the SI cross section is of the form
\begin{align}
\sigma_{\mathrm{SI}} \sim \frac{m_p^4m_\chi^2}{m_{\tilde f}^8}.
\end{align}
 The SI direct detection cross section for Majorana DM is therefore strongly suppressed for most of the parameter space and will not lead to significant bounds. Although the spin-dependent (SD) cross section is not suppressed, the current bounds from SD direct detection experiments \cite{Aprile:2013doa} are too loose to lead to significant constraints on the DM and mediator masses. For models involving both left and right-handed squarks, introducing a left-right mixing generates a dimension six effective operator contributing to the spin-independent DD cross section. Obtaining this left-right mixing in a natural way without inducing flavor mixing, however, leads to mixing angles that are strongly suppressed for light squark flavors. We show in the Appendix that introducing these small mixing angles does not significantly increase the spin-independent DD cross sections for models with Majorana DM.
\section{Results}
\label{Sec: sec5}
\noindent
In this section, we give the results 
when DM couples to the mediators: $\tilde{f} = $\{$\tilde u_R$, $\tilde d_R$, $\tilde q_L$, $\tilde L_L$, $\tilde L_R$$\}$.  For each case, we perform a uniform scan over
\begin{align}
\lambda_i \in [0,5] && m_\chi, m_{\tilde{f}_i} \in [0, 100] \,\,\mathrm{TeV} \label{eq:couplingscan} 
\end{align}
where $i = L, R$.  We compute the DM relic density and direct detection cross section using MicrOmegas \cite{Belanger:2013oya} and generate the model files using FeynRules \cite{Alloul:2013bka} interfaced with CalcHEP \cite{Belyaev:2012qa}. We require the relic density to be less than the value measured by the Planck satellite~\cite{Ade:2013lta} plus $3\, \sigma$, i.e.,
\begin{align}
h^2 \Omega_c \le 0.1226.
\label{Eq: relic}
\end{align}
We require the points in the model to pass the LUX bounds \cite{Akerib:2013tjd}.  We also check whether or not the surviving points are within the projected reach of  XENON1T \cite{Aprile:2012nq}.  

\subsection{Results for Colored Flavor Triplet Effective Model}
\label{Sec: sec5coloredtripletresults}
\begin{figure}
\centering
\includegraphics[width=0.5\linewidth]{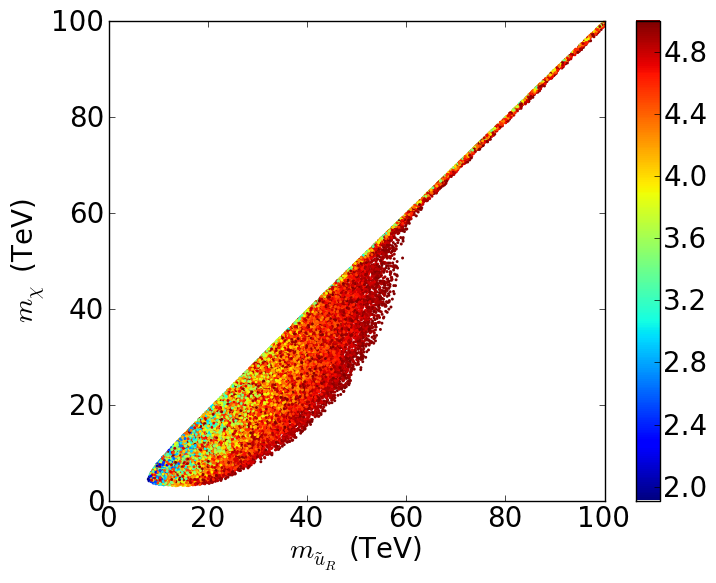}\\
\includegraphics[width=0.5\linewidth]{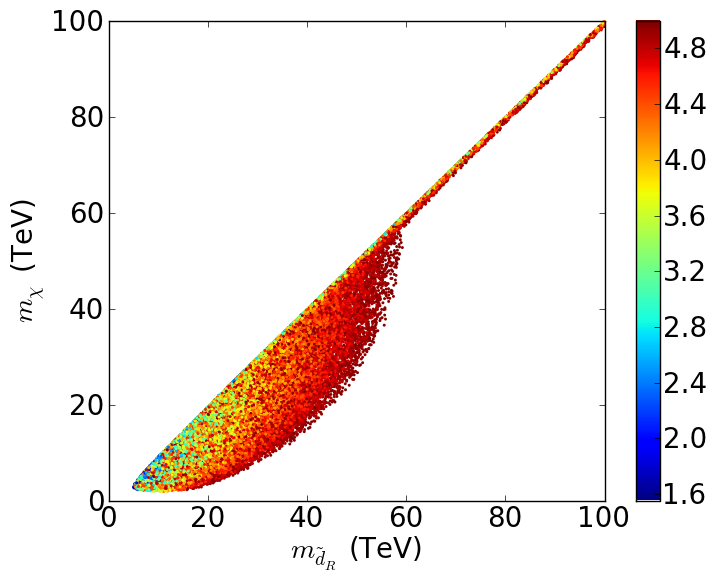}\\
\includegraphics[width=0.5\linewidth]{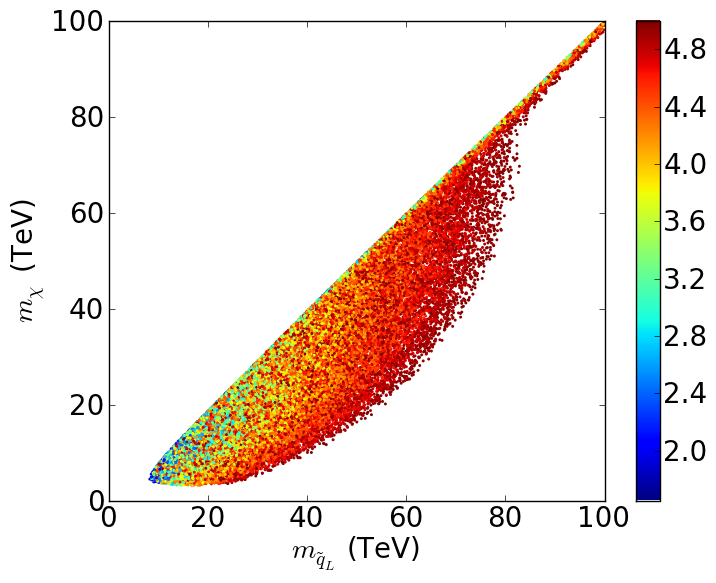}
\caption{\label{msqDirac}Points that pass the unitarity, relic density and direct detection constraints in the $(m_\chi, m_{\tilde f})$ space for $\tilde u_R$ (top plot), $\tilde d_R$ (middle plot) and $\tilde q_L$ (bottom plot) with Dirac DM. The narrow region along the diagonal is the coannihilation region.  The colors denote the magnitude of the coupling of $\lambda_i$.  See equations~\leqn{Eq: lag3} and~\leqn{eq:couplingscan}.  (Color online.)}
\end{figure}

\begin{figure}
\centering
\includegraphics[width=0.5\linewidth]{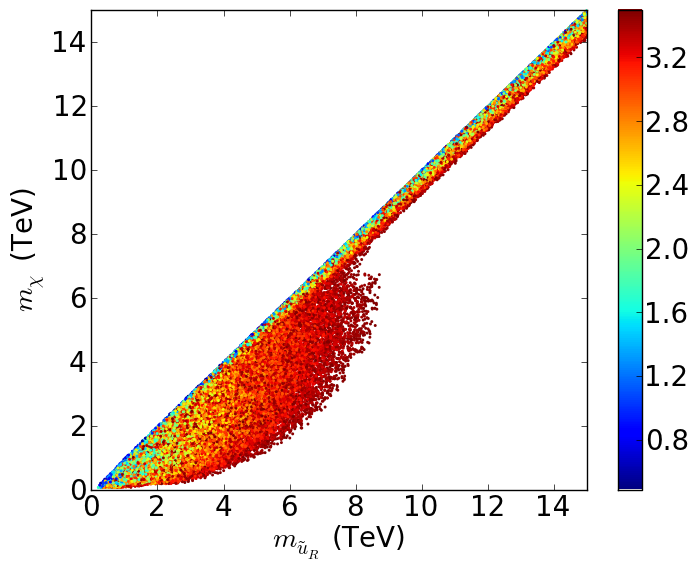} \\
\includegraphics[width=0.5\linewidth]{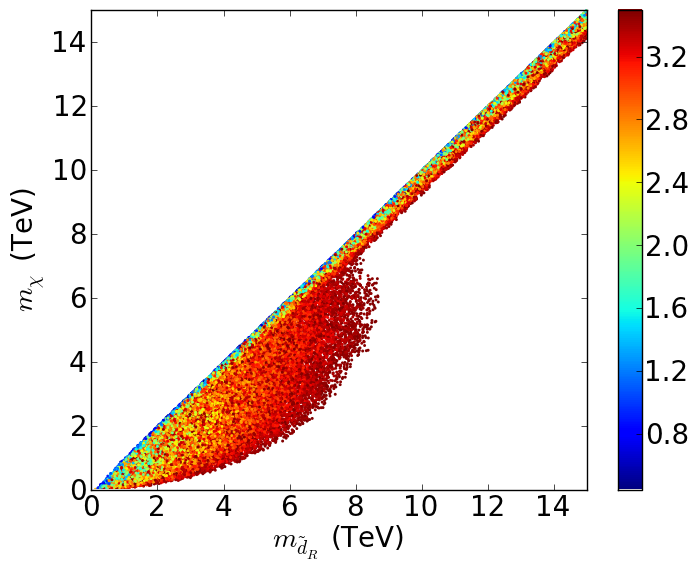}\\
\includegraphics[width=0.5\linewidth]{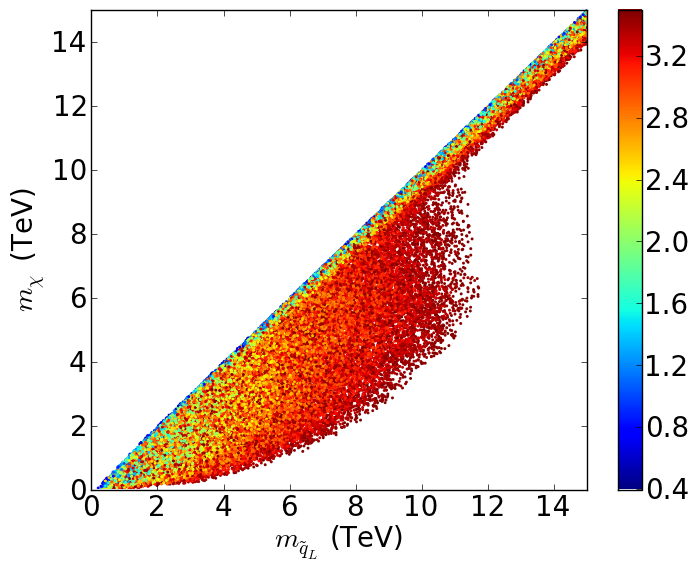}
\caption{\label{msqMaj}Points that pass the unitarity, relic density and direct detection constraints in the $(m_\chi, m_{\tilde f})$ space for $\tilde u_R$ (top plot), $\tilde d_R$ (middle plot) and $\tilde q_L$ (bottom plot) with Majorana DM. The narrow region along the diagonal is the coannihilation region.  The colors denote the magnitude of the coupling of $\lambda_i$.  See equations~\leqn{Eq: lag3} and~\leqn{eq:couplingscan}.  (Color online.)}
\end{figure}

\begin{figure}
\centering
\includegraphics[width=0.5\linewidth]{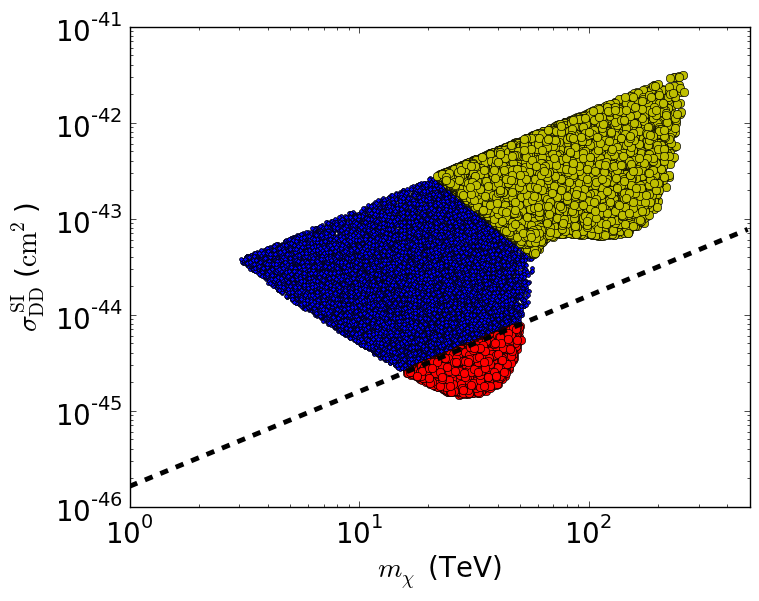}\\
\includegraphics[width=0.5\linewidth]{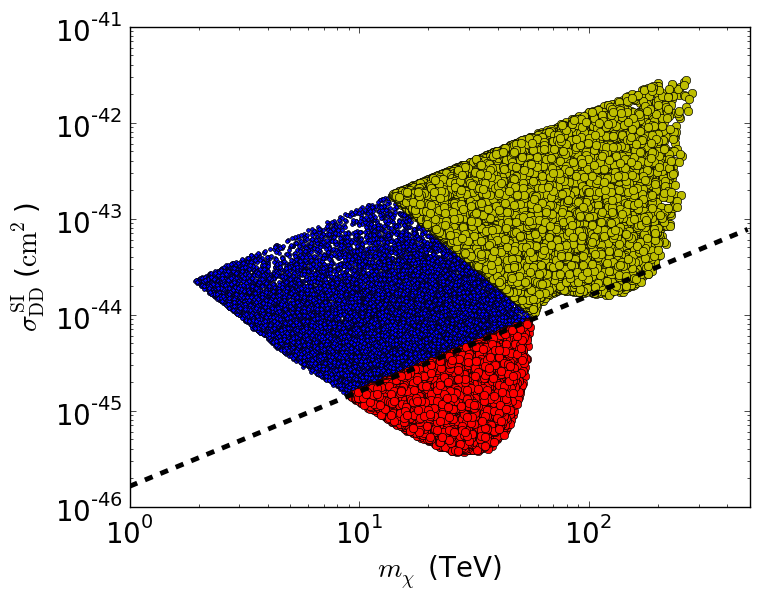}\\
\includegraphics[width=0.5\linewidth]{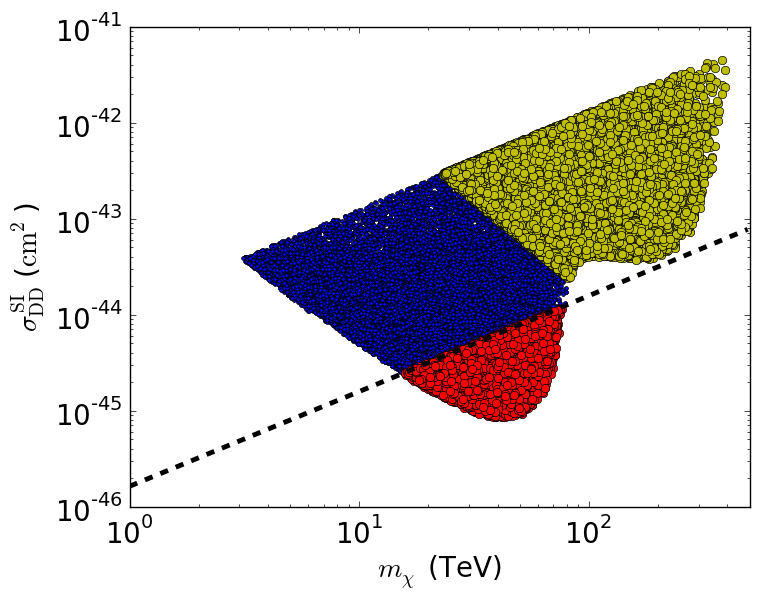}
\caption{\label{DD} Spin-independent direct detection cross section vs $m_\chi$ for the points that pass the unitarity, relic density and direct detection constraints for $\tilde u_R$, $\tilde d_R$ (top left and right) and $\tilde q_L$ (bottom) with Dirac DM. For each parameter point, we rescale the direct detection cross section by the ratio of the relic density of $\chi$ over the total DM relic density shown in equation~\leqn{Eq: relic}. The yellow points are the points in the coannihilation region while the red point are the points that are outside the reach of XENON1T.  See equation~\leqn{eq:coannihil}. The black dashed line represents the projected XENON1T bound extrapolated for large DM masses.}
\end{figure}

\begin{figure}
\centering
\includegraphics[width=0.43\linewidth]{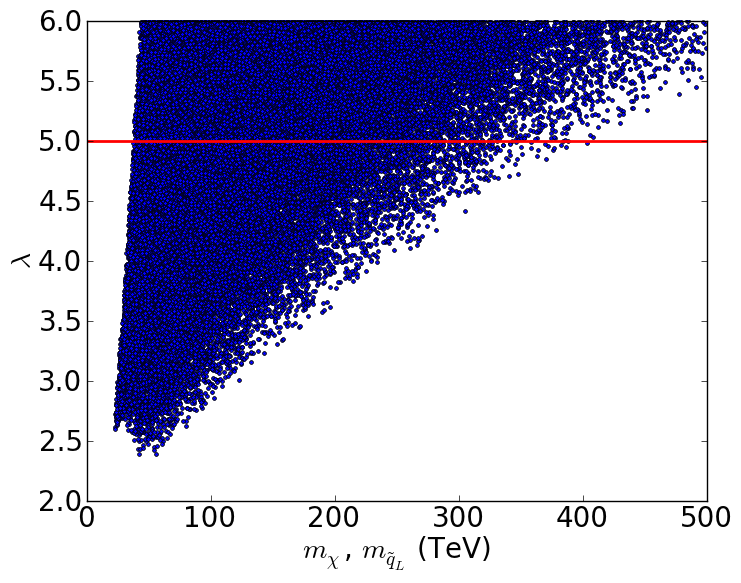}
\includegraphics[width=0.42\linewidth]{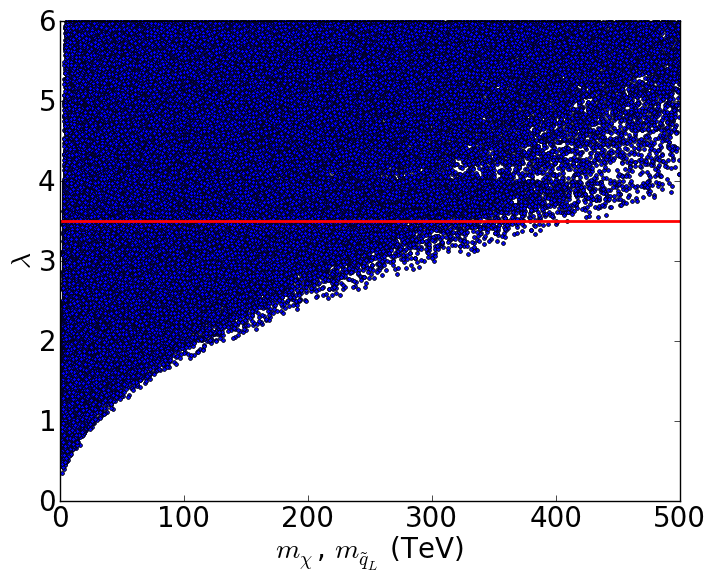}
\includegraphics[width=0.43\linewidth]{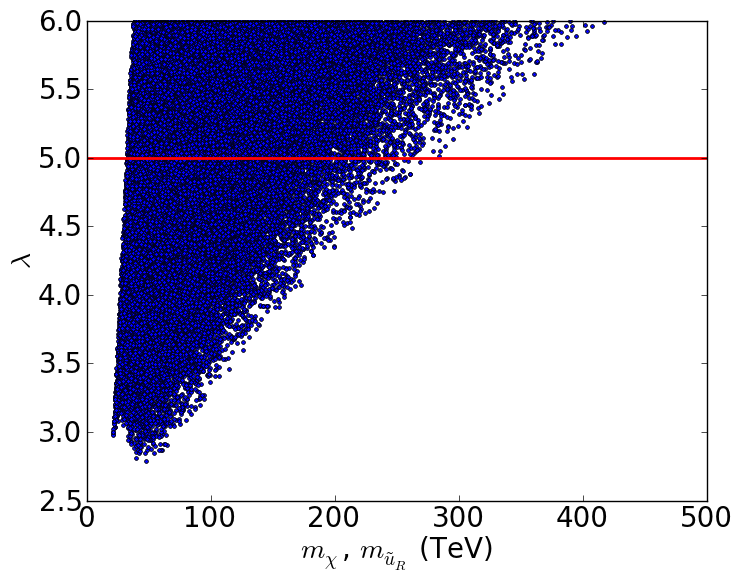}
\includegraphics[width=0.42\linewidth]{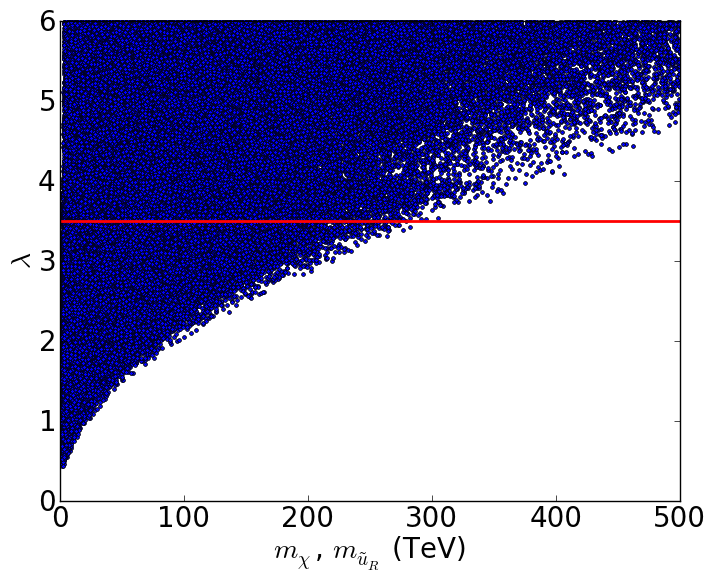}
\includegraphics[width=0.43\linewidth]{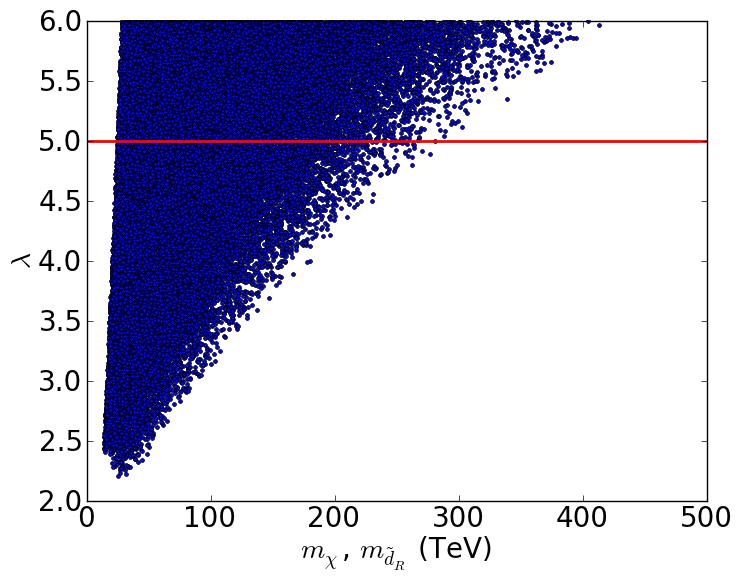}
\includegraphics[width=0.42\linewidth]{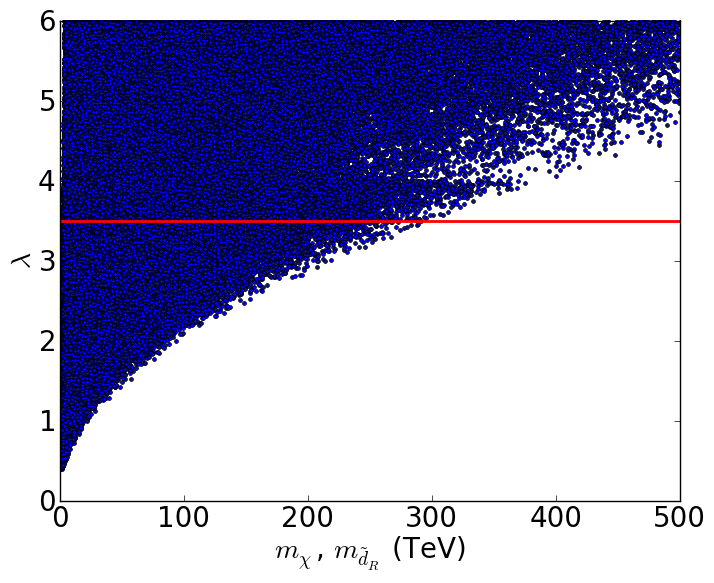}
\caption{\label{coann}Points in the coannihilation region that pass the unitarity, relic and DD constraints for $\tilde{q}_L$ (top), $\tilde{d}_R$ (middle), and $\tilde{u}_R$ (bottom) for Dirac (left) and Majorana (right) DM. The horizontal line is the unitarity bound on the coupling $\lambda$.}
\end{figure}

\noindent
Figure~\ref{msqDirac} shows the points that pass the unitarity, relic density and direct detection constraints in the $(m_\chi, m_{\tilde{f}_i})$ space for $\tilde u_R$, $\tilde d_R$ and $\tilde q_L$ with Dirac DM. The narrow region along the diagonal is the coannihilation region, in which the relic density bounds are considerably loosened. Outside this region, the bounds on the DM and squark masses are around $60\,\,\mathrm{TeV}$ for $\tilde u_R$ and $\tilde d_L$ and around $80\,\,\mathrm{TeV}$ for $\tilde q_L$ for which there exist six annihilation channels instead of three. Note that these upper bounds vary as $\sqrt{N_{\mathrm{flavors}}\,N_c}$, as explained in Section~\ref{Sec: sec4}.
Figure~\ref{msqMaj} shows the points that pass the unitarity, relic density and direct detection constraints in the $(m_\chi, m_{\tilde f})$ space for $\tilde u_R$ (top plot), $\tilde d_R$ (middle plot) and $\tilde q_L$ (lower plot) with Majorana DM. As expected, the relic density constraint leads to tighter bounds than in the Dirac case. The bounds outside the coannihilation region are now around $10$ TeV for all squark families.
Figure~\ref{DD} shows the direct detection cross section vs the DM mass for the points  that pass the unitarity, relic density and the LUX direct detection constraints. For each parameter point, we rescale the direct detection cross section by the ratio of the relic density of $\chi$ over the total DM relic density shown in equation~\leqn{Eq: relic}. The points in red represent the points not generated by co-annihilation that are outside the projected reach of XENON1T.  The yellow points are solely in the co-annihilation region.  There were select by requiring
\begin{align}
\frac{\left|m_\chi - m_{\tilde f}\right|}{m_\chi} < 5\%. \label{eq:coannihil}
\end{align}
Direct detection experiments are extremely sensitive to the squark parameter space for dirac DM and the future experiments will be able to exclude almost all of our models, even in the case of co-annihilation.  
\newline
\newline
We also consider co-annihilating points in parameter space that satisfy equation~\leqn{eq:coannihil}.  These points %
can surpass the well known bound from Griest and Kamionkowski~\cite{Griest:1989wd} since it is not valid in case of coannihilation.  Figure~\ref{coann} shows the points that pass the unitarity, relic density and direct detection constraints in the $(\lambda, m_\chi)$ plane for $\tilde u_R$, $\tilde{d}_R$ and $\tilde{q}_L$.  Both Dirac and Majorana DM are considered.  We find that these bounds are around $300$ TeV for the squarks.

\subsection{Results for Stop Effective Model}
\noindent
We plot the points %
that pass the unitarity, relic density and direct detection constraints in the $(m_\chi, m_{\tilde t_R})$ space for RH stops, for Dirac and Majorana DM in Appendix~\ref{Sec: appendixB}, Figure~\ref{unitarityrelicplots}. For this model, we find bounds of $35$ TeV for Dirac DM and $5$ TeV for Majorana DM.  Since we consider only one flavor here, the bounds on the DM and stop masses are tighter than the squark bounds by a factor of about $\sqrt{3}$. 
Since the top component in the nucleons is largely subdominant, no significant DD bound can be obtained. 

\subsection{Results for Colored Singlet Flavor Triplet Effective Model}
\noindent
We plot the %
points that pass the unitarity, relic density and direct detection constraints in the $(m_\chi, m_{\tilde f})$ space for $\tilde e_R$ and $\tilde l_L$ with Dirac and Majorana DM in Appendix~\ref{Sec: appendixB}.  Here the bounds are tighter than for the squarks since the slepton annihilation is not enhanced by the color factor. The mass bounds are therefore decreased by a factor of $\sqrt{3}$.  For Dirac DM, we find a bound of $35$ TeV for $\tilde e_R$ and $50$ TeV for $\tilde l_L$.  For Majorana DM, we find a bound $5$ and $9$ TeV for $\tilde e_R$ and $\tilde l_L$, respectively.  Also Appendix~\ref{Sec: appendixB}, Figure~\ref{coann2} has the slepton results for the co-annihiliating points which satisfy equation~\leqn{eq:coannihil}.
%
%

\subsection{Comments on Collider Constraints}
\begin{figure}
\centering
\vspace{-1.25in}
\includegraphics[width=0.7\linewidth]{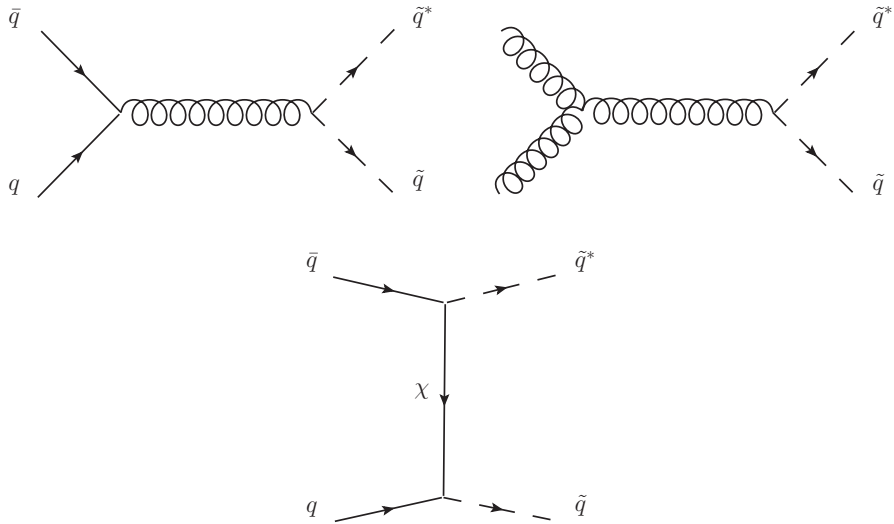}
\vspace{-1.5in}
\caption{\label{Fig: squarkprod} Important squark pair production modes. For the large coupling $\lambda$ considered here, the $t$-channel diagram at the bottom of the figure can contribute significantly to the squark pair production cross section.  Thus, if these models are responsible for the measured relic abundance, the sensitivity of a $100$ TeV hadron collider can be greatly enhanced.}
\end{figure}

\noindent
Majorana DM models as well as all models with only stops or sleptons are not significantly constrained by direct detection. However, both DM and sfermions can be searched for at colliders. In SUSY, the dominant production mechanism for squarks is pair-production with an intermediate gluon. In our study, $\lambda$ is required by relic density constraints to be large for heavy DM, and $t$-channel squark production through the diagram shown in Figure~\ref{Fig: squarkprod} starts becoming significant.  Bounds on the $\tilde u_R$, $\tilde d_R$ and $\tilde q_L$ models using the $8$ TeV ATLAS and CMS data have been derived in \cite{DiFranzo:2013vra}. These bounds are about a few hundreds of $\mathrm{GeV}$ and exclude only a tiny fraction of the parameter space considered here. A $100$ TeV hadron collider, however, could exclude squarks up to a few $\mathrm{TeV}$, as shown in \cite{Cohen:2013xda,Cohen:2014hxa}. Such a collider could exclude a sizeable part of the parameter space, especially for Majorana DM. For DM heavier than a few $\mathrm{TeV}$, the relic density requirement constrains the masses of the DM and the squarks to be close to each other, and traditional squark searches start losing their power. Due to the size of the coupling $\lambda$, monojet searches are a serious competitor to squark searches in these kinematically challenging regions.
\section{Conclusions}
\label{Sec: Conclusion}
\noindent
Understanding the nature of dark matter is one of the most compelling question in particle physics.  %
Dark matter that was once in thermal equilibrium with the SM is very well motivated.  In this work we focus on scenarios where the thermal dark matter annihilates via new charged and/or colored scalars (sfermions).  %
We have shown here that combining unitarity arguments with relic density constraints allows to derive upper bounds on both the masses of the DM and of the sfermion mediators. Using flavor 
arguments, we consider various possible sfermion configurations, notably, degenerate sfermion flavor triplets and models where the stop is much lighter than the other new particles. For Dirac DM, the upper bounds on the DM and sfermion masses  vary between about $35$ TeV for stops and right-handed sleptons to $80$ TeV for left-handed squark doublets outside the coannihilation region. The bounds for Majorana DM are much tighter, varying from $5$ TeV for stops and right-handed sleptons to about $10$ TeV for left-handed squark doublets. We also consider co-annihilation regions, in which the Griest and Kamionkowski bound does not apply. In these regions, we derive upper bounds on the DM mass as high as $400$ TeV.  The upper bounds derived in this work strongly depend on the number of scalar mediators driving the DM annihilation. This work, however, spans a wide range of possible configurations and, within the MFV framework, can be readily adapted to models involving different combinations of sfermions.
\newline
\newline
Exploiting the tension between unitarity and relic density constraints on DM and sfermions not only allows us to derive upper bounds on the scales of new physics, but also to define goals for future experiments. For Dirac DM, in particular, future direct detection experiments such as XENON1T could probe most of the parameter space in models with light-flavor squarks, leaving only extremely narrow regions viable. For Majorana DM or models with stops or sleptons, however, collider searches and indirect detection will provide the best constraints on new physics. For Majorana DM, which is associated with particularly tight bounds, a future $100\mathrm{TeV}$ collider would be able to probe most of the parameter space for squark models. Since relic density constraints require the DM and sfermion masses to be close to each other, however, developing and improving compressed spectrum searches will be crucial to get a better understanding of scalar mediator models.
\noindent

\Acknowledgements
We thank J.~Hewett, A.~E.~Nelson, M.~Peskin, T.~Rizzo and D.~Zeppenfeld for useful discussions.  This work is supported by the US Department of Energy, contract DE-ACO2-76SF00515. DW is also supported in part by a grant from the Ford Foundation via the National Academies of the Sciences as well as the National Science Foundation under Grants No. NSF PHY11-25915 and NSF-PHY-0705682. SEH is supported by the Cluster of Excellence Precision Physics, Fundamental Interactions and Structure of Matter (PRISMA-EXC 1098).

\appendix

\section{Squark Flavor Triplets with Left-Right Mixing}
\label{Sec: appendix}
\noindent
If we allow for a non-trivial coupling between the sfermion mediators and the SM Higgs that is analogous to the A-terms in SUSY, we can generate  left-right (L-R) mixing after electroweak symmetry breaking (EWSB).  This can be important for models with flavor triplet squarks and Majorana DM.  In this case, the spin-independent direct detection cross section is suppressed due to its twist-2 origin.  The direct detection constraints could in principle become stronger if the left and right-handed squarks are allowed to mix. This scenario has been most recently studied in the context of SUSY for sneutrino-mediated DM annihilation~\cite{Kelso:2014qja}. To explore this possibility, we consider a model with one left-handed squark doublet $\tilde q_L$ and one right-handed up-type squark $\tilde u_R$, with the right-handed down-type squark $\tilde d_R$ decoupled. Since mixing between different squark flavors is strongly constrained by kaon oscillations \cite{Altmannshofer:2013lfa}, we only allow mixing between left and right-handed squarks of the same flavor. Such a scenario can be obtained naturally by requiring the minimum flavor violation (MFV)~\cite{Chivukula:1987py,D'Ambrosio:2002ex}, and requiring the left-right squark  mixing terms in the Lagrangian to be proportional to the masses of the corresponding quarks.
 The mass and interaction terms for the squarks are then taken to be
\begin{align}
V_{\mathrm{mass}} &= \frac{1}{2} m_{\tilde{q}_L}^2 \tilde{q}_L^*\tilde{q}_L + \frac{1}{2} m_{\tilde{u}_R}^2 \tilde{u}_R^*\tilde{u}_R\\
V_{\mathrm{int}} &= \lambda_L\tilde{q}_L^* \bar\chi P_L q + \lambda_R \tilde{u}_R \bar\chi P_R u + \mathrm{h.c.}\\
V_{\mathrm{Yukawa}} &= A_{LR} Y_u \tilde{q}_L^* H \tilde{u}_R + \mathrm{h.c.}
\label{Eq: Lmixing}
\end{align} 
\begin{figure}
\centering
\includegraphics[width=0.6\linewidth]{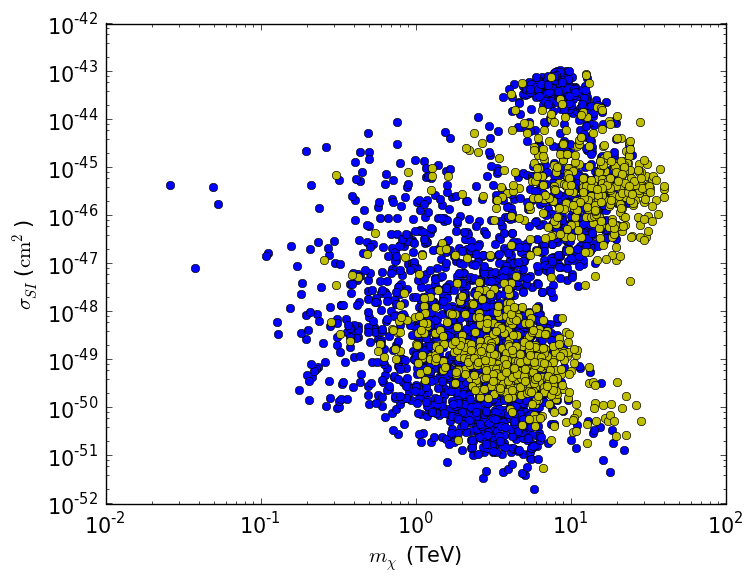}
\caption{\label{LRmixed}$\sigma_{\mathrm{SI}}$ vs $m_\chi$ for points with $\theta_3 < 0.1$ (blue) and $\theta_3 > \frac{\pi}{4} - 0.1$ (green). The points displayed here have passed the unitarity, relic density and LUX constraints.}
\end{figure}
where $Y_u$ is the matrix of the SM up-type Yukawa couplings, $H$ is the SM Higgs, $A_{LR}$ is an A-term-like parameter with dimensions of energy, and the squark mass matrices $m_{\tilde{q}_L}$ and $m_{\tilde{u}_R}$ are proportional to the identity. The Yukawa potential in equation \leqn{Eq: Lmixing} induces a mass mixing between the up-type component of $\tilde{q}_L$ and $\tilde{u}_R$ after EWSB. The mass eigenstates $\tilde u_1$ and $\tilde u_2$ are then such that
\begin{align}
\tilde u_L &= \cos\theta\,\tilde u_1  + \sin\theta\,\tilde u_2 \\
\tilde u_R &= -\sin\theta\,\tilde u_1  + \cos\theta\, \tilde u_2 .
\end{align}
\begin{figure}
\centering
\includegraphics[width=0.45\linewidth]{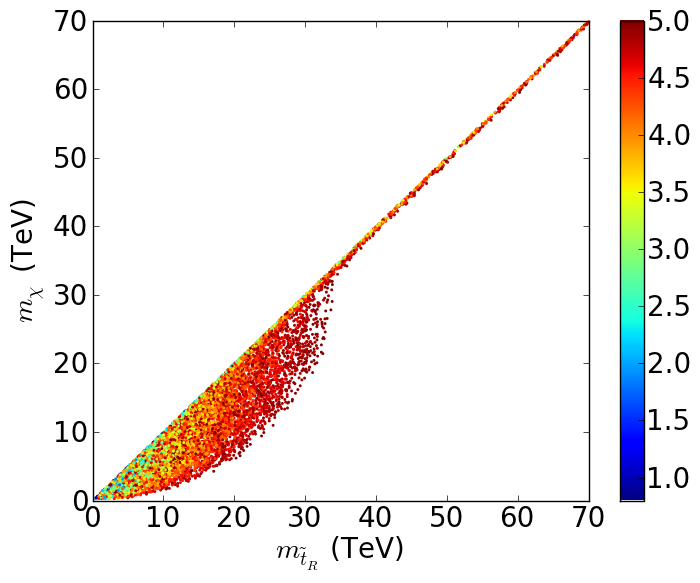}
\includegraphics[width=0.45\linewidth]{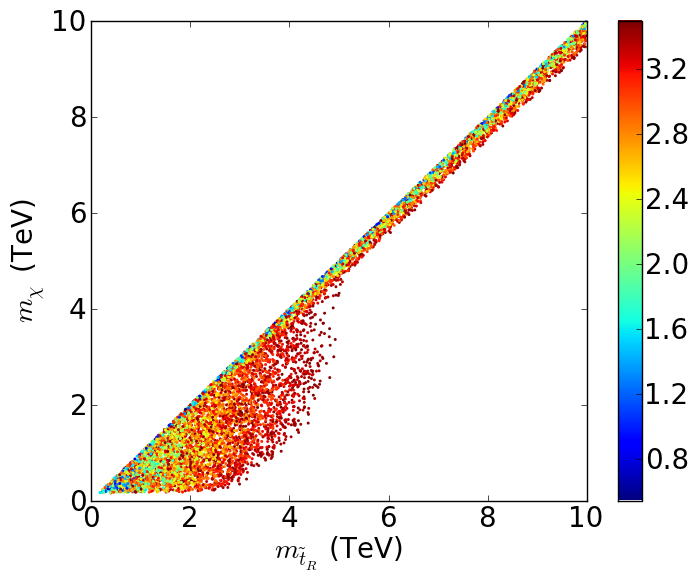} \\
\includegraphics[width=0.45\linewidth]{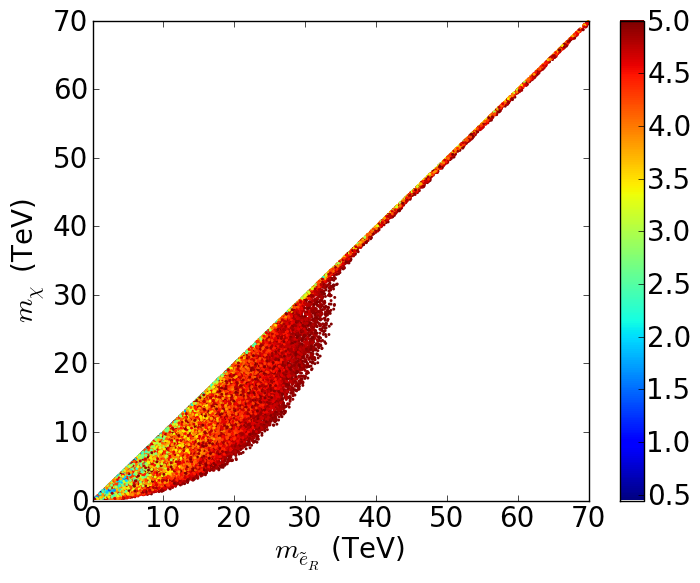}
\includegraphics[width=0.45\linewidth]{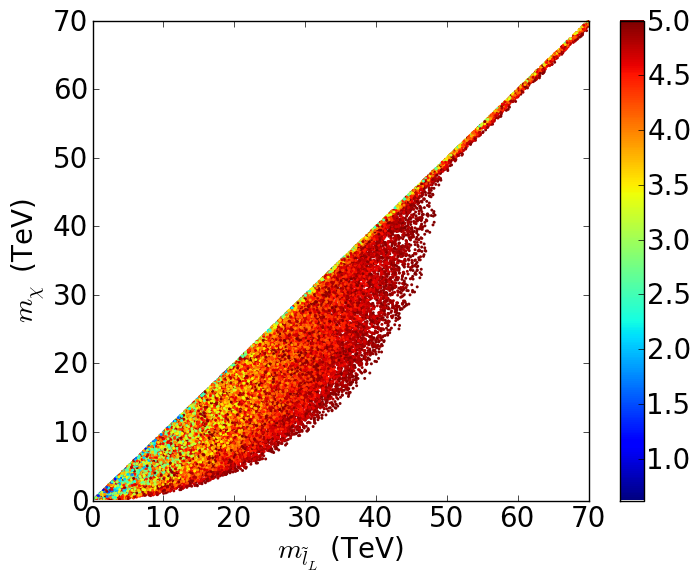} \\
\includegraphics[width=0.45\linewidth]{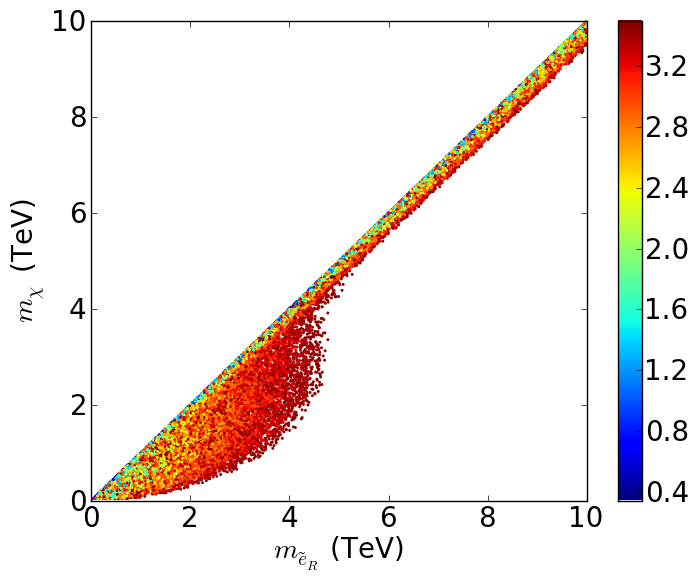}
\includegraphics[width=0.45\linewidth]{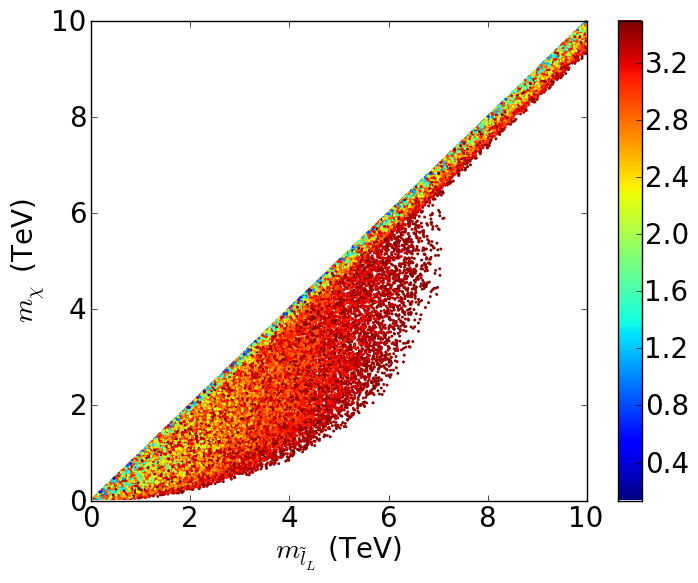}
\caption{\label{unitarityrelicplots} Top plots: Points that pass the unitarity, relic density and direct detection constraints in the $(m_\chi, m_{\tilde t_R})$ pace for $\tilde t_R$ with Dirac (left) and Majorana (right) DM.  Middle plots: Points that pass the unitarity, relic density and direct detection constraints in the $(m_\chi, m_{\tilde f})$ space for $\tilde e_R$ (left) and $\tilde l_L$ (right) with Dirac DM.  Bottom plots:  Points that pass the unitarity, relic density and direct detection constraints in the $(m_\chi, m_{\tilde f})$ space for $\tilde e_R$ (left) and $\tilde l_L$ (right) with Majorana DM.  %
For each plot, the narrow regions along the diagonal is the co-annihilation region.}
\end{figure}
Since the L-R mixing term from the squark-Higgs interaction is proportional to the SM Yukawa couplings, the mixing angle $\theta$ is flavor-dependent, with the stops mixing much more strongly than the charm and up squarks. Since we have decoupled $\tilde d_R$, the LH down squarks do not mix with anything. The DM interaction terms with $\tilde u_1$, $\tilde u_2$ and $\tilde d_L$ are then
 \begin{align}
V_\mathrm{int} &= \tilde{u}_1^*\, \bar\chi\,\Bigl(\lambda_L \cos\theta\, P_L - \lambda_R\sin\theta P_R\Bigr) u   + \tilde{u}_2^*\, \bar \chi 
\Bigl(\lambda_L \sin\theta\, P_L + \lambda_R\cos\theta P_R\Bigr)\,u + \tilde d_L \bar\chi P_L d+ \mathrm{h.c.}
\label{Eq: lagmix} 
 \end{align}
Unlike the previous effective scenarios, the resulting effective model has six free parameters,
\begin{align}
\left\{m_\chi, m_{\tilde q_L}, m_{\tilde u_R}, A_{LR}, \lambda_L, \lambda_R\right\}.
\end{align}
For Majorana DM, introducing left-right mixing generates dimension six effective operators that contribute to the spin-independent Dark Matter direct detection cross section. For a given squark flavor $i$, the lowest order mixing contribution to the SI cross section is of the form \cite{Kelso:2014qja}
\begin{align}
\sigma_{SI} &\propto \lambda_L^2\lambda_R^2 \sin^2 2\theta_i\left[\frac{1}{m_{\tilde{u}_1^i}^2 - m_\chi^2} - \frac{1}{m_{\tilde{u}_2^i}^2 - m_\chi^2}\right]^2.
\end{align}
 Due to the hierarchy of the SM Yukawa couplings, sizeable mixing will occur only in the stop sector. For light flavor squarks, the enhancement of the DD cross section due to introducing left-right mixing is therefore expected to be negligible. Fig.~\ref{LRmixed} shows the points that pass the unitarity, relic and LUX constraints in the $\left(m_\chi, \sigma_{\mathrm{SI}}\right)$ plane. The blue points are points for which the stop left-right mixing is low
\begin{align}
\theta_3 \le 0.1
\end{align}
while the yellow points are points for which the stop left-right mixing is large
\begin{align}
\theta_3 > \frac{\pi}{4} - 0.1.
\end{align}
We removed the points belonging to the coannihilation region by requiring
\begin{align}
\frac{\left| m_\chi - m_{\tilde{u}_1^3}\right|}{m_\chi} > 10\%, 
\end{align}
$m_{\tilde{u}_1}^3$ being the mass of the lightest stop, which is also the lightest squark in our model.
As expected, introducing left-right mixing does not lead to a significant global improvement of the spin-independent DD cross section for Majorana DM.
\begin{figure}[!t]
\centering
\includegraphics[width=0.45\linewidth]{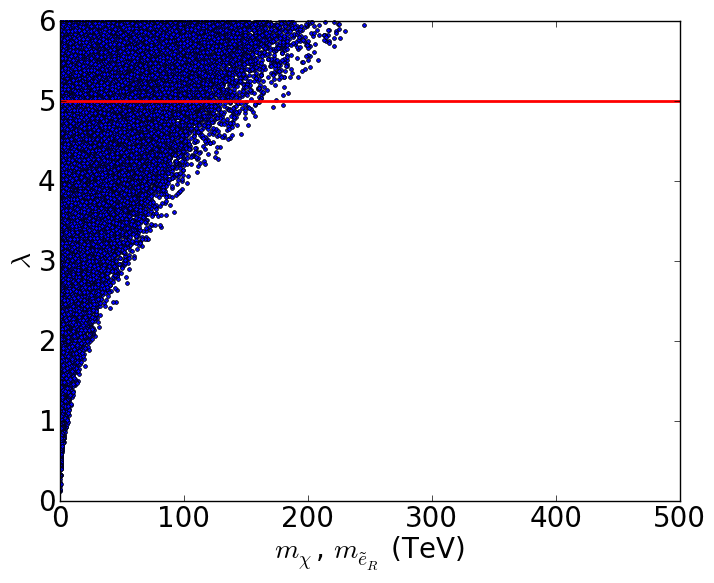}
\includegraphics[width=0.45\linewidth]{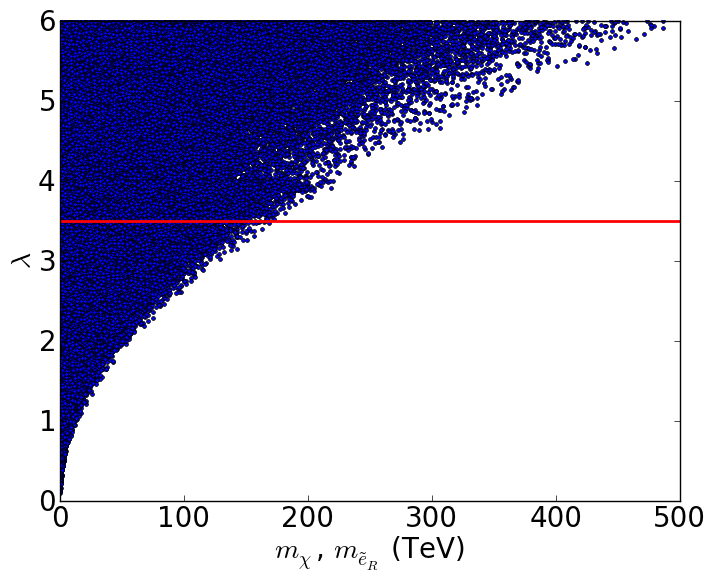}
\includegraphics[width=0.45\linewidth]{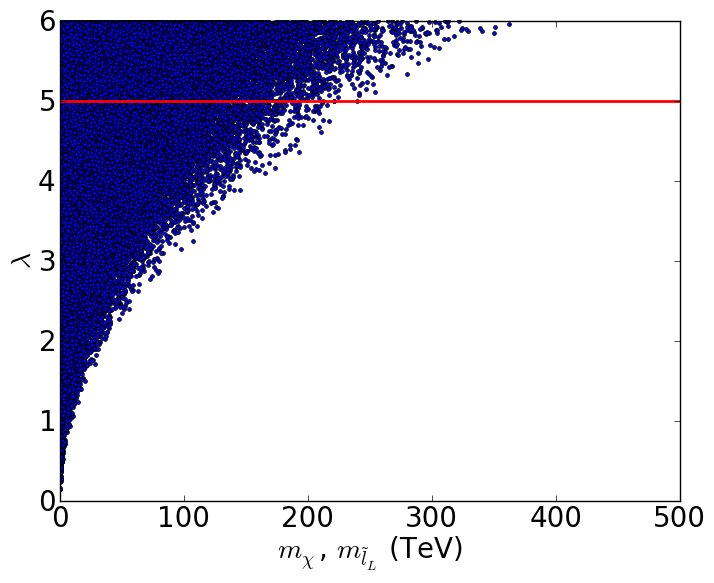}
\includegraphics[width=0.45\linewidth]{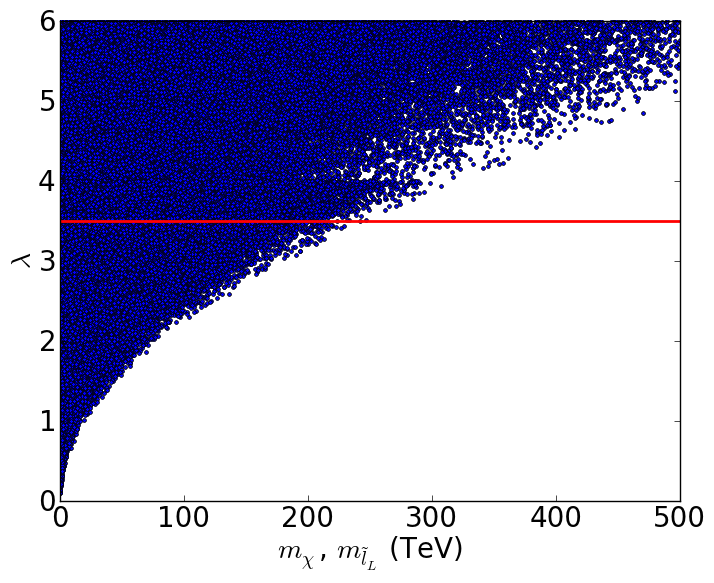}
\caption{\label{coann2}Points in the coannihilation region that pass the unitarity, relic and DD constraints for $\tilde{e}_R$ (top) and $\tilde{l_L}$ (bottom) for Dirac (left) and Majorana (right) DM. The horizontal line is the unitarity bound on the coupling $\lambda$.}
\end{figure}

\section{Additional Results for Various Effective Models} 
\label{Sec: appendixB}
\noindent
Considering the effective field theory with only stops and dark matter, we plots all of the points that pass the unitarity, relic density and direct detection constraints in Figure~\ref{unitarityrelicplots}.  The plots are in the $(m_\chi, m_{\tilde t_R})$ space for RH stops, for Dirac and Majorana DM.  Because there is only one squark flavor, the bounds on the DM and stop masses are tighter than the squark bounds by a factor of about $\sqrt{3}$. 
\newline
\newline
%
Figure~\ref{unitarityrelicplots} also shows the unitarity, relic density and direct detection constraints for the effective field theory with flavor triplet sleptons.  Again, the plot is in the $(m_\chi, m_{\tilde f})$ space for $\tilde e_R$ and $\tilde l_L$ with Dirac and Marjorana DM.  Again, the bounds are tighter than for the squarks bounds shown in the text since the slepton annihilation is not enhanced by the color factor. The mass bounds are therefore decreased by a factor of $\sqrt{3}$.  Since $N_\mathrm{flavor} \,N_c$ is the same for the $\tilde e_R$ model (reminder, three flavors) and the stop-only model, the bounds on the DM and squark masses are the same in both cases.
%
\newline
\newline
Finally, we consider the parameter space points in the co-annhilation regime for the slepton effective theories.  Figures~\ref{coann} and \ref{coann2} shows the  points that pass the unitarity, relic density and direct detection constraints in the $(\lambda, m_\chi)$ plane for  $\tilde{l}_L$ and $\tilde{e}_R$.  We consider both Dirac and Majorana DM. As emphasized before, since Griest and Kamionkowski~\cite{Griest:1989wd} is not valid in case of coannihilation, these bounds can be well beyond $120$ TeV. We find $200$ TeV for RH sleptons for both Dirac and Majorana DM.

\end{document}